\def\be{\begin{equation}}
\def\ee{\end{equation}}
\def\bea{\begin{eqnarray}}
\def\eea{\end{eqnarray}}

\documentclass[doublecol]{epl2} 
\newcommand{\overlim}[1]{{\buildrel{#1}\over\rightarrow\;}}

\title{Quasi-Integrability in Supersymmetric Sine-Gordon Models}

\author{K. Abhinav\inst{1} \and P. Guha\inst{1}}

\institute{                    
  \inst{1} SN Bose National Centre for Basic Sciences - JD Block, Sector III, Salt Lake, Kolkata 700106,  India}
\pacs{02.30.Ik}{Integrable systems}
\pacs{11.30.Pb}{Supersymmetry}
\pacs{05.45.Yv}{Solitons}

\abstract{The deformed supersymmetric sine-Gordon model, obtained through known deformation of the
corresponding potential, is found to be {\it quasi-integrable}, like its non-supersymmetric counterpart,
which was observed earlier. The system expectedly possesses finite number of conserved quantities, leaving-out
an infinite number of non-conserved anomalous charges. The quasi-integrability of this supersymmetric
model heavily rely on the boundary conditions of the potential, otherwise rendered to be completely
non-integrable. Moreover, interesting additional algebraic structures appear, absent in the
non-supersymmetric counterparts.}

\begin{document}

\maketitle

\section{Introduction}

The integrability of finite dimensional dynamical systems is related to the fact that the system possesses as many constant
of motion as its number of degrees of freedom and it is described by the Liouville-Arnold theorem.
There is no such unique definition to describe integrability for dynamical systems having infinite degrees of freedom or partial 
differential equations (PDEs). The notion of integrability is described via various ways, for example, bi-Hamiltonian theory,
method of inverse scattering transformation, Lax pair formalism etc.. The integrable PDEs appearing in field theory
are best studied via Lax pair method or zero curvature equation. These are deemed integrable if they contain 
infinitely many conserved quantities, which are responsible for the stability of corresponding soliton solutions \cite{Das}.
In particular, these constants of motion uniquely define the dynamics
of the system, rendering the later to be completely solvable. The sine-Gordon (SG) model, in one space and
one time (1+1) dimensions, is one such system that further incorporates semi-classical solitonic 
solutions, which are physically realizable, representing high degree of symmetry. The latter property,
in turn, corresponds to the infinitely many conserved quantities. Such solitonic solutions of integrable
models correspond to the zero-curvature condition \cite{Z1,Z2}, containing the connections that
constitute the Lax pair, which linearize the corresponding non-linear system. 

Real physical systems do not posses infinite degrees of freedom, and thus, a corresponding
field-theoretical model cannot be integrable in principle. However, such systems do posses solitonic 
states, which are very similar to those of integrable models like SG. Therefore, study of physical continuous
systems as slightly deformed integrable models is of conceptual interest. Recently, it was shown that 
the SG model can be deformed as an approximate system, leading to a finite number of conserved quantities
\cite{f1,2}. An almost flat connection, leading to an anomalous zero-curvature condition was obtained,
rendering the system {\it quasi-integrable}. it was numerically shown \cite{f1,f3} that the system behaves
almost like the integrable SG model for single and non-interacting multi-soliton states, but differ
considerably in presence of scattering. This hints at integrable models being asymptotic limits of 
physical systems, which can be modeled by their quasi-integrable counterparts.

On the other hand, the SG model admits a supersymmetric extension, which can be embedded
in the $N=2$ superspace \cite{SSG0}. This model also is integrable \cite{SSG1,1} and can consistently be quantized
\cite{QSSG1,QSSG2}. Their integrability, through linearization by introducing proper Lax pair, leads
to the super SG (SSG) equation as a consistency condition. Such a model contains fermionic components 
of the Majorana variety, which is of great interest in current material physics. Latter it was shown
that the SSG admits two independent ways of linearization, corresponding to two possible construction 
of the Lax pair \cite{Siddiq}. However, these two representations were found to be gauge-equivalent.
The model was further shown to incorporate multi-soliton solutions \cite{Siddiq1}. As the quasi-integrable
models are closer to physical systems, a corresponding deformation of SSG model can potentially lead 
to the realization of the same as an limiting case. This is further important from the pure algebraic
sense, as the generalization to the superspace can expectedly induce richer mathematical structure.

In the present work, we attempt to construct the quasi-integrable SSG (QISSG) model. It
is found that the generalization to the superspace indeed extends the corresponding group structure
beyond the usual $sl(2)$ loop algebra of the SG model \cite{f1,2}. However, the quasi-integrability,
obtained through gauge transformation of the anomalous curvature condition, is retained and a finite
number of conserved charges appear, with the asymptotic correspondence to the SSG generalized to the 
fermionic `boundary'. In the following, we briefly introduce the nomenclature of SSG model, as well as that of
quasi-integrability, followed by our generalization of the latter to the SSG model. We summarize the
results at the end.
 
\section{Basic Concepts}

\subsection{Super Sine-Gordon Model}
The sine-Gordon (SG) model \cite{Das}, defined through the dynamical equation,

\be
\partial_+\partial_-\phi(x,t)=\sin\phi(x,t),\label{p1}
\ee
is an integrable one, containing infinitely many conserved quantities. This system is realized in 
1+1 dimensions, and it is convenient to cast it in terms of light-cone
coordinates, $x_\pm=(1/2)\left(x\pm t\right)$. The respective partial derivatives appear in
the equation of motion (Eq. \ref{p1}) above. The integrability of this model incorporates
an $sl(2)$ loop algebra \cite{sl2}, which automatically leads to infinite number of conserved 
quantities upon linearization through introduction of the Lax pair, by implementing the 
zero-curvature condition \cite{Das}. The Lax pair $A_\pm$, that constitute the curvature
function $F_{+-}=\partial_+A_--\partial_-A_++[A_+,A_-]$, are $2\times2$ matrices, corresponding to 
the two-dimensional coordinate space. As a result, the $sl(2)$ generators also have a $2\times2$
irreducible representation.

A supersymmetric generalization of the SG model was achieved \cite{1}, by
extending it to the $N=2$ superspace through introduction of additional Grassmann variables
$\theta_{1,2}$, which absolutely anti-commute: $\{\theta_i\theta_j\}=0,\quad i,j=1,2$. The resultant 
supersymmetric sine-Gordon (SSG) equation has the form,

\be
D_+D_-\Phi=i\sin\Phi.\label{1}
\ee
Here the super-differential operators and the superspace bosonic scalar function, respectively,
are defined as \cite{1},

\bea
D_\pm&=&\mp\frac{\partial}{\partial\theta_{2,1}}+i\theta_{2,1}\frac{\partial}{\partial x_\pm},~\{D_+,D_-\}=0,\nonumber\\
\Phi(x,t,\theta_1,\theta_2)&=&\varphi(x,t)+i\left(\theta_1\psi_2(x,t)-\theta_2\psi_1(x,t)\right)\nonumber\\
&+&i\theta_1\theta_2F(x,t),\label{2}
\eea
In the above, $\psi_{1,2}$ are odd and $\varphi(x,t), F(x,t)$ are even functions of
space-time, making $\Phi(x,t,\theta_1,\theta_2)$ a scalar (bosonic) function in the super-space.

The linearization of the SSG model is achieved \cite{1}, by identifying 
the corresponding Lax pair, which are now $3\times3$ matrices in the superspace. Further, 
they were found to implement integrability through the zero-curvature condition. It was later
shown \cite{Siddiq} that the linear representation is not unique, and two such forms, related
to each other by a super gauge transformation and to the super B\"acklund transformation of
the equation.  

\subsection{Quasi-Integrable Models}
Certain non-integrable theories share physical results similar to the integrable ones. Namely,
solitonic solutions of certain respective models, belonging to this distinct class, have
similar properties \cite{f1}. The same is observed in 2+1 dimensions, for the Ward-modified chiral
model and baby Skyrme model having many potentials \cite{2}. Therefore, it has recently been
attempted to identify such non-integrable models as quasi-integrable ones \cite{f1,f2}, in 
order to realize such models as parametric generalization of the integrable counterparts.
Such generalization is complemented by anomaly functions $P_n$, defined through \cite{2},
$$\frac{dQ_n}{dt}=P_n,\quad n\in {\cal Z},$$
wherein the quantities $Q_n$ are conserved for vanishing $P_n$, a limit that leads to the 
corresponding integrable model. 
When one considers $P_n$ for two soliton configurations in these theories, the corresponding anomaly has the intriguing property
$\int_{-\infty}^{\infty} dt\,P_n = 0$, which implies that the charges are asymptotically conserved:
$Q_n (t = −\infty) = Q_n (t = +\infty)$. If the model possesses breather-like field configurations we obtain
$Q_n (t) = Q_n (t + T )$. On the other hand, for different configurations, like 
multi-soliton solutions, the anomalies $P_n$ were found to be non-vanishing, but they display
interesting boundary properties, having topological nature \cite{2}.  

As an concrete example, a deformation of the SG potential is proposed as a
quasi-integrable system, leading to a quasi zero-curvature condition: $F_{+-}\neq0$. a suitable
gauge transformation then leads to quasi-conservation equations, with only a few anomaly 
functions ($P_n$s) vanishing. We generalize the construction of quasi-integrable SG model
to the supersymmetric one, or a quasi-integrable supersymmetric sine-Gordon (QISSG) equation,
in the following.

\section{The Supersymmetric Model}
To begin with, we consider the $N=2$ supersymmetric SG model \cite{1}, that yields the Lax
pair, 

\bea
{\cal A}_+&=&\frac{i}{2\sqrt{2\lambda}}\frac{dV(\Phi)}{d\Phi}F_1+\frac{1}{2\sqrt{2\lambda}}V(\Phi)B_1,\nonumber\\
{\cal A}_-&=&\frac{1}{\sqrt{2\lambda}}B_{-1}-\frac{i}{2}D_-\Phi F_0.\label{3}
\eea
Here, the $3\times3$ generators have the forms, 

\bea
F_1&=&2\sqrt{\lambda}\left( \begin{array}{ccc} 0 & 0 & -i \\ 0 & 0 & 0 \\ -i & 0 & 0 \end{array} \right),\nonumber\\
B_1&=&2\sqrt{\lambda}\left( \begin{array}{ccc} 0 & 0 & 0 \\ 0 & 0 & -1 \\ 0 & -1 & 0 \end{array} \right),\nonumber\\
B_{-1}&=&2\sqrt{\lambda}\left( \begin{array}{ccc} 0 & 0 & 0 \\ 0 & 0 & -1 \\ 0 & 1 & 0 \end{array} \right),\nonumber\\
F_0&=&2\left( \begin{array}{ccc} 0 & i & 0 \\ -i & 0 & 0 \\ 0 & 0 & 0 \end{array} \right).\label{4}
\eea
This generators act on a vector-space, defined by the supersymetric generalization of the
$sl(2)$ loop algebra, identified in Ref. \cite{2}. The super Lax pair leads to the
super-curvature tensor, 

\bea
{\cal F}_{+-}&=&D_+{\cal A}_--D_-{\cal A}_++\left\{{\cal A}_+,{\cal A}_-\right\}\nonumber\\
&\equiv&\frac{i}{2\sqrt{2\lambda}}{\cal X}(\Phi)F_1-\frac{i}{2}E(\Phi)F_0,\label{5}
\eea 
where, 

\bea
{\cal X}(\Phi)&=&\left(\frac{d^2V(\Phi)}{d\Phi^2}+V(\Phi)\right)D_-\Phi,\nonumber\\
E(\Phi)&=&D_+D_-\Phi+\frac{dV(\Phi)}{d\Phi};\label{6}\\
V(\Phi)&=&i\cos\Phi,\nonumber
\eea
and $\lambda$ is the spectral parameter. The Euler function $E(\Phi)$ vanishes on-shell (Eq. \ref{1}),
and ${\cal X}(\Phi)$ is the generator of anomaly, which basically is the differential form satisfied 
by the sine-Gordon potential function $V(\Phi)$, and therefore vanishes. However, this is
{\it not} the case for quasi-integrable deformation of the same. The deformation is done by
introducing a small parameter $\varepsilon\ll 1$, which leads to the quasi-integrable potential,

\be
V(\Phi,\varepsilon)=i\left[1-\frac{32\tan^2\left(\frac{\Phi}{4}\right)}{(2+\varepsilon)^2}\left(1-\left\vert\sin\frac{\Phi}{4}\right\vert^{2+\varepsilon}\right)^2\right],\label{7}
\ee
that reduces to $V(\Phi)$ for $\varepsilon=0$. More importantly, it renders ${\cal X}(\Phi,\varepsilon)\neq0$, 
leading to quasi-integrability and thus, non-vanishing curvature ${\cal F}_{+-}(\Phi,\varepsilon)$.

\subsection{The Super-$sl(2)$ Loop Algebra}
As the generators of Lax pair in the superspace corresponding to the super-SG model has
been identified, the generators of the corresponding $N=2$ super-$sl(2)$ loop algebra can 
be defined through the hierarchy, 
\bea
B_{2n+1}&=&\lambda^n\left(\tau_++\lambda\tau_-\right),\quad F_{2n+1}=\lambda^n\left(\tau_+-\lambda\tau_-\right),\nonumber\\
F_{2n}&=&2\lambda^n\tau_3.\label{8}
\eea
The generators of the above algebra have the construction,  

\bea
\tau_+&=&\frac{1}{2}\left(B_1+F_1\right),\qquad\tau_-=\frac{1}{2\lambda}\left(B_1-F_1\right),\nonumber\\
\tau_3&=&\frac{1}{2}F_0,\label{9}
\eea
leading to the $sl(2)$ algebra,

\be
[\tau_+,\tau_-]=2\tau_3,\qquad[\tau_3,\tau_\pm]=\pm\tau_\pm.\label{10}
\ee
The above algebraic structure can be utilized to rotate the curvature function in Eq. \ref{5},
in the $sl(2)$ vector space, thereby extracting-out the role of anomaly function ${\cal X}(\Phi)$.

\subsection{Gauge Transformation}
The definition of the curvature function designates the Lax pair ${\cal A}_\pm$ as gauge
fields, allowing for a gauge transformation of the kind \cite{f1,2},

\bea
&&{\cal A}_\pm\rightarrow{\cal B}_\pm=G{\cal A}_\pm G^{-1}-D_\pm GG^{-1},\nonumber\\
&&G=\exp\left[\sum_{n=1}^\infty\zeta_nF_n\right].\label{2}
\eea
The choice of the transformation generator $G$ utilizes the $sl(2)$ algebraic structure,
thereby making this transformation essentially a rotation in the space of the super-$sl(2)$ loop
algebra. The form of $G$ enables the utilization of the Baker-Campbell-Hausdorff (BCH) formula,

$$e^XY_\pm e^{-X}=Y_\pm+[X,Y_\pm]+\frac{1}{2!}[X,[X,Y_\pm]]+\cdots,$$
to evaluate the first term on the LHS of Eq. \ref{2}, where it can be identified that,

\be
X=\sum_{n=1}^\infty\zeta_nF_n\qquad\&\qquad Y_\pm={\cal A}_\pm.\nonumber
\ee
The rotated Lax component, ${\cal B}_+$ is evaluated to begin with, as a choice, which is not unique. 
This is what is expected from the notion of gauge-invariance. However, the choice of representation
for the particular gauge group can result in analytical simplicity. In the present case, the additional presence 
of supersymmetry makes this choice effectively crucial, and prohibits the matrix $B_{-1}$, in Eq. \ref{4} from being a
semi-simple element of the $sl(2)$ algebra in Eq. \ref{10}. It can owe to the fermionic domain of the supersymmetric
structure, incorporating its non-commutative nature in an indirect manner. Therefore, from
the representation chosen in Eqs. \ref{3} and \ref{4}, we choose to evaluate ${\cal B}_+$ first,
unlike the case of the non-supersymmetric system of Ref. \cite{2}, where the negative component was chosen. In any case,
the components will have the form,

\be
{\cal B}_\pm\equiv e^XY_\pm e^{-X}-\sum_nD_\pm\zeta_nF_n.\label{p3}
\ee
The first four non-trivial terms of the BCH expansion are of the form,

\bea
&&[X,Y_+]\nonumber\\
&&=\alpha(\Phi)\zeta_{2m}B_{2m+1}+\beta(\Phi)\Big\{\zeta_{2m}F_{2m+1}+\zeta_{2m-1}F_{2m}\Big\},\nonumber\\
&&\frac{1}{2!}[X,[X,Y_+]]\nonumber\\
&&=\alpha(\Phi)\zeta_{2m}\Big\{\zeta_{2p}F_{2(m+p)+1}+\zeta_{2p-1}F_{2(m+p)}\Big\}\nonumber\\
&&+\beta(\Phi)\Big\{\zeta_{2m}\zeta_{2p}B_{2(m+p)+1}-\zeta_{2m-1}\zeta_{2p-1}B_{2(m+p)-1}\Big\},\nonumber\\
&&\frac{1}{3!}[X,[X,[X,Y_+]]]\nonumber\\
&&=\frac{2}{3}\Big[\alpha(\Phi)\zeta_{2m}\Big\{\zeta_{2p}\zeta_{2q}B_{2(m+p+q)+1}\nonumber\\
&&\qquad\quad-\zeta_{2p-1}\zeta_{2q-1}B_{2(m+p+q)-1}\Big\}+\beta(\Phi)\zeta_{2m}\zeta_{2p}\nonumber\\
&&\qquad\times\Big\{\zeta_{2q}F_{2(m+p+q)+1}+\zeta_{2q-1}F_{2(m+p+q)}\Big\}\nonumber\\
&&\qquad-\beta(\Phi)\zeta_{2m-1}\zeta_{2p-1}\Big\{\zeta_{2q}F_{2(m+p+q)-1}\nonumber\\
&&\qquad\qquad\qquad\qquad\qquad+\zeta_{2q-1}F_{2(m+p+q-1)}\Big\}\Big],\nonumber\\
&&\frac{1}{4!}[X,[X,[X,[X,Y_+]]]]\nonumber\\
&&=\frac{1}{3}\Big[\alpha(\Phi)\zeta_{2m}\zeta_{2p}\Big\{\zeta_{2q}\Big(\zeta_{2r}F_{2(m+p+q+r)+1}\nonumber\\
&&\qquad+\zeta_{2r-1}F_{2(m+p+q+r)}\Big)+\zeta_{2q-1}\Big(\zeta_{2r}F_{2(m+p+q+r)-1}\nonumber\\
&&\qquad+\zeta_{2r-1}F_{2(m+p+q+r-1)}\Big)\Big\}\nonumber\\
&&+\beta(\Phi)\zeta_{2m}\zeta_{2p}\Big\{\zeta_{2q}\zeta_{2r}B_{2(m+p+q+r)+1}\nonumber\\
&&\qquad\qquad\qquad-\zeta_{2q-1}\zeta_{2r-1}B_{2(m+p+q+r)-1}\Big\}\nonumber\\
&&-\beta(\Phi)\zeta_{2m-1}\zeta_{2p-1}\Big\{\zeta_{2q}\zeta_{2r}B_{2(m+p+q+r)-1}\nonumber\\
&&\qquad\qquad\qquad\quad-\zeta_{2q-1}\zeta_{2r-1}B_{2(m+p+q+r)-3}\Big\}\Big];\label{40}
\eea
where,
\bea
\alpha(\Phi)&=&\frac{i}{\sqrt{2\lambda}}\frac{dV(\Phi)}{d\Phi}, \beta(\Phi)=\frac{1}{\sqrt{2\lambda}}V(\Phi),~\&\nonumber\\
(m,p,q,r)&=&1,2,3,\cdots\nonumber
\eea
The gauge-fixing is implemented through the condition that the coefficients of $F_n$, for all $n$ in
Eq. \ref{p3}, should vanish. This leads to the following consistency conditions:  
\vskip 0.25cm
\noindent For $F_1$: $\qquad D_+\zeta_1=\frac{1}{2}\alpha(\Phi)$,
\vskip 0.25cm
\noindent For $F_2$: $\qquad D_+\zeta_2=\beta(\Phi)\zeta_1$,
\vskip 0.25cm
\noindent For $F_3$: $\qquad D_+\zeta_3=\beta(\Phi)\zeta_2$,
\vskip 0.25cm
\noindent For $F_4$: $\qquad D_+\zeta_4=\beta(\Phi)\left(\zeta_3-\frac{2}{3}\zeta_1^3\right)+\alpha(\Phi)\zeta_2\zeta_1$,
\vskip 0.2cm
\noindent and so on. These relate the parameters with the variables of the system. 
The above equations have to be integrated to obtain $\zeta_n$'s, including $\alpha(\Phi)$ and $\beta(\Phi)$
which, in general, can require non-trivial boundary conditions, both in $\zeta$-parameter space and $\Phi$-field space.
Such behavior can be attributed to the supersymmetric nature of the present system, as discussed before.
As a result, the
gauge-fixed field component has the form,

\be
{\cal B}_+\equiv\sum_{n=0}^\infty b_+^{(2n+1)}B_{2n+1}.\label{7}
\ee
with various coefficients $b_+^{(2n+1)}$:

\bea
&&b_+^{(1)}=\frac{1}{2}\beta(\Phi),\nonumber\\
&&b_+^{(3)}=\alpha(\Phi)\zeta_2-\beta(\Phi)\zeta_1^2,\nonumber\\
&&b_+^{(5)}=\alpha(\Phi)\left\{\zeta_2-\frac{2}{3}\zeta_2\zeta_1^2\right\}\nonumber\\
&&\qquad\quad+\beta(\Phi)\left\{\zeta_2^2-2\zeta_3\zeta_1+\frac{1}{3}\zeta_1^4\right\},\nonumber\\
&&\cdots\nonumber\\
&&\cdots\nonumber\\
&&\cdots\label{8}
\eea
It is to be noted that, in evaluating ${\cal B}_+$, the equation of motion has not been used, {\it i.
e.}, the Euler function $E(\Phi,\varepsilon)$, corresponding to the deformed potential Eq. \ref{7},
has not been set equal to zero. Therefore, the above results are {\it off-shell}. To determine the exact form of the
coefficients $b_+^{(2n+1)}$, the terms $\zeta_n$ are to be evaluated, for which the super-differential
equations are obtained as consistency conditions. The necessary boundary conditions
for the same is provided by the equation of motion, which will be used in evaluating the other Lax
component ${\cal B}_-$. For this purpose, we adopt the definitions, 

\bea
{\cal A}_-&=&\frac{\kappa}{2}B_{-1}+\frac{\gamma(\Phi)}{2}F_0;\nonumber\\
\kappa&=&\sqrt{\frac{2}{\lambda}},\qquad \gamma(\Phi)=-iD_-\Phi.\label{N1}
\eea
As in case of ${\cal B}_+$, the first three non-trivial BCH contributions have the form,

\bea
&&[X,Y_-]\nonumber\\
&&=\kappa\Big\{\zeta_{2m}{\bar F}_{2m+1}+\zeta_{2m-1}{\bar F}_{2m}\Big\}-\gamma(\Phi)\zeta_{2m-1}B_{2m-1},\nonumber\\
&&\frac{1}{2!}[X,[X,Y_-]]\nonumber\\
&&=\kappa\Big[\zeta_{2m}\Big\{\zeta_{2p}B^{m+p}_{-1}+2\zeta_{2p-1}{\tilde F}_{2(m+p)-1}\Big\}\nonumber\\
&&\qquad+\zeta_{2m}\Big\{2\zeta_{2p}{\tilde F}_{2(m+p)}-\zeta_{2p-1}B^{m+p-1}_{-1}\Big\}\Big]\nonumber\\
&&-\gamma(\Phi)\zeta_{2m-1}\Big[\zeta_{2p}F_{2(m+p)-1}+\zeta_{2p-1}F_{2(m+p-1)}\Big],\nonumber\\
&&\frac{1}{3!}[X,[X,[X,Y_-]]]\nonumber\\
&&=\frac{2}{3}\kappa\Big[\zeta_{2m}\Big\{\zeta_{2p}\Big(\zeta_{2q}{\bar F}_{2(m+p+q)+1}+\zeta_{2q-1}{\bar F}_{2(m+p+q)}\Big)\nonumber\\
&&\quad+\zeta_{2p-1}\Big(\zeta_{2q}{\bar F}_{2(m+p+q-1)}-4\zeta_{2q-1}{\bar F}_{2(m+p+q)-3}\Big)\Big\}\nonumber\\
&&\quad+\zeta_{2m-1}\Big\{4\zeta_{2p}\Big(\zeta_{2q}{\bar F}_{2(m+p+q)}-\zeta_{2q-1}{\bar F}_{2(m+p+q)-1}\Big)\nonumber\\
&&\quad-\zeta_{2p-1}\Big(\zeta_{2q}{\bar F}_{2(m+p+q)-1}+\zeta_{2q-1}{\bar F}_{2(m+p+q-1)}\Big)\Big\}\Big]\nonumber\\
&&-\frac{2}{3}\gamma(\Phi)\zeta_{2m-1}\Big[\zeta_{2p}\zeta_{2q}B_{2(m+p+q)-1}\nonumber\\
&&\qquad\qquad\qquad-\zeta_{2p-1}\zeta_{2q-1}B_{2(m+p+q)-3}\Big];\label{9}
\eea
wherein the following additional matrices appear:

\bea
{\bar F}_{2m}&=&\lambda^m{\bar F}_0,\nonumber\\
{\bar F}_0&=&\frac{1}{2\lambda}\Big[F_1,B_{-1}\Big]=2\left( \begin{array}{ccc} 0 & -i & 0 \\ -i & 0 & 0 \\ 0 & 0 & 0 \end{array} \right),\nonumber\\
{\bar F}_{2m+1}&=&\lambda^m{\bar F}_1,\nonumber\\
{\bar F}_1&=&\frac{1}{2}\Big[F_0,B_{-1}\Big]=2\sqrt{\lambda}\left( \begin{array}{ccc} 0 & 0 & -i \\ 0 & 0 & 0 \\ i & 0 & 0 \end{array} \right),\nonumber\\
{\tilde F}_{2m}&=&\lambda^m{\tilde F}_0,\nonumber\\
{\tilde F}_0&=&\frac{1}{4}\Big[F_0,{\bar F}_0\Big]=2\left( \begin{array}{ccc} 1 & 0 & 0 \\ 0 & -1 & 0 \\ 0 & 0 & 0 \end{array} \right),\nonumber\\
{\tilde F}_{2m+1}&=&\lambda^m{\tilde F}_1,\nonumber\\
{\tilde F}_1&=&\frac{1}{4\lambda}\Big[F_1,{\bar F}_1\Big]=2\left( \begin{array}{ccc} 1 & 0 & 0 \\ 0 & 0 & 0 \\ 0 & 0 & -1 \end{array} \right),\nonumber\\
B^n_{-1}&=&\lambda^nB_{-1},\quad\&\quad(m,p,q,r)=1,2,3,\cdots\label{9a}
\eea
The above matrices act in the space of the super-space multiplates, as are the original matrices 
in Eq. \ref{4}, but they construct enclosed matrix vector subspaces, leading to a twisted loop algebra,
of the form obtained in Ref. \cite{TLA}. To see this, one notes the following partially close algebra:

\bea
\Big[F_0,{\bar F}_1\Big]&=&2B_{-1},\quad\Big[F_1,{\bar F}_0\Big]=-2B_{-1},\nonumber\\
\Big[F_0,{\tilde F}_1\Big]&=&2{\bar F}_0,\quad\Big[F_1,{\tilde F}_1\Big]=-4{\bar F}_1,\nonumber\\
\Big[F_0,{\tilde F}_0\Big]&=&4{\bar F}_0,\quad\Big[F_1,{\tilde F}_0\Big]=-4{\bar F}_1.\label{N2}
\eea
From the above algebra, it is easy to identify the rotations, in the space of $\left[B_{-1},{\bar F}_{0,1}\right]$, 

\bea
&&B_{-1}\overlim{F_0}{\bar F}_1\overlim{F_0}B_{-1},\nonumber\\
&&B_{-1}\overlim{F_1}{\bar F}_0\overlim{F_1}B_{-1};\label{N10}
\eea
generated by $F_{0,1}$ independently. However, rotation in a larger subspace $\left[B_{-1},{\bar F}_{0,1},{\tilde F}_{0,1}\right]$, is also identified as,
\bea
&&B_{-1}\overlim{F_0}{\bar F}_1\overlim{F_1}{\tilde F}_1\overlim{F_0}{\bar F}_0\overlim{F_1}B_{-1},\nonumber\\
&&B_{-1}\overlim{F_1}{\bar F}_0\overlim{F_0}{\tilde F}_0\overlim{F_1}{\bar F}_1\overlim{F_0}B_{-1};\label{11}
\eea
due to the action of same pair of operators $F_{0,1}$, however, now acting successively. Such extension
to the matrix vector space can be considered as quasi-integrable generalization of the emergent twisted
loop algebra.
These additional algebraic structures further ensure the uniqueness of the SSG model, as inherent
supersymmetry does not allow $B_{-1}$ to be a part of the $sl(2)$ algebra, in the present representation of Eq. \ref{4}.
Eqs. \ref{N2}, \ref{N10} and \ref{11} are unique to the present supersymmetric quasi-integrable model, absent in the
non-supersymmetric counterparts \cite{2,f3}. This is an alternate signature of the fermionic sector non-commutativity,
following the apparent non-locality of parameters $\zeta_n$ and charges $b^{(2n+1)}_+$, as discussed before.
The final expression for the gauge field ${\cal B}_-$ has the form,

\bea
&&{\cal B}_-\equiv\sum_{n=0}^\infty\Big[b_{-1}^{(n)}B_{-1}^{(n)}+b_-^{(2n+1)}B_{(2n+1)}\nonumber\\
&&\qquad\qquad\qquad+f^{(n)}F_{n}+{\bar f}^{(n)}{\bar F}_{n}+{\tilde f}^{(n)}{\tilde F}_{n}\Big],\label{12}
\eea
with non-zero coefficients, 

\bea
&&b_{-1}^{(0)}=\frac{1}{2}\kappa,\quad b_{-1}^{(1)}=-\kappa\zeta_1^2,\quad b_{-1}^{(2)}=\kappa\zeta_2^2,\cdots;\nonumber\\
&&b_-^{(1)}=-\gamma(\Phi)\zeta_1,\quad b_-^{(3)}=-\gamma(\Phi)\Big(\zeta_3-\frac{2}{3}\zeta_1^3\Big),\nonumber\\
&&b_-^{(5)}=-\gamma(\Phi)\Big(\zeta_5+\frac{2}{3}\zeta_1\zeta_2^2-2\zeta_3\zeta_1^2\Big)+\cdots;\nonumber\\
&&f^{(0)}=\frac{1}{2}\gamma(\Phi),\quad f^{(1)}=-D_-\zeta_1,\quad f^{(2)}=-D_-\zeta_2-\gamma(\Phi)\zeta_1^2,\nonumber\\
&&f^{(3)}=-D_-\zeta_3-\gamma(\Phi)\zeta_2\zeta_1,\cdots;\nonumber\\
&&{\bar f}^{(2)}=\kappa\zeta_1\quad{\bar f}^{(3)}=\kappa\Big(\zeta_2-\frac{8}{3}\zeta_2\zeta_1^2\Big),\nonumber\\
&&{\bar f}^{(4)}=\kappa\Big(\zeta_3+\frac{2}{3}\zeta_2^2\zeta_1-\frac{2}{3}\zeta_1^3\Big)+\cdots;\nonumber\\
&&{\tilde f}^{(3)}=2\kappa\zeta_2\zeta_1,\quad{\tilde f}^{(4)}=2\kappa\zeta_2\zeta_1+\cdots.\nonumber\\
&&\cdots\nonumber\\
&&\cdots\nonumber\\
&&\cdots\label{13}
\eea
In evaluating the exact expressions for the above coefficients, the equation of motion,
$(dV/d\varphi)+D_+D_-\varphi=0$, has been utilized. This fact, as will be seen, enables us to determine the original
coefficients $\zeta_n$ of the gauge transformation, for the integrable part of the system.

\subsection{The quasi-Integrability}
The anomaly $X$ vanishes for integrable models. In the present non-integrable one, due to its presence,
the anomalous coefficients $f^{(n)}, {\bar f}^{(n)} \& {\tilde f}^{(n)}$ survive.
It is observed that these coefficients vanish for $X=0$. This allows for the definition of the
quasi-continuity expressions,

\be
\Gamma^{{2n+1}}:=D_+b_-^{(2n+1)}-D_-b_+^{(2n+1)},\quad n=0,1,2,\cdots\label{14}
\ee
from Eqs. \ref{7} and \ref{12}. The first of the above expressions vanishes for
$\zeta_1=-\frac{i}{\sqrt{2\lambda}}D_-\Phi$, through the equation of motion. This is exactly the
supersymmetric generalization of the result in \cite{2}. Further, the first of the charge-densities,

\be
\Sigma^{(2n+1)}:=\frac{1}{2}\left(b_+^{(2n+1)}-b_-^{(2n+1)}\right),\quad n=0,1,2,\cdots,\label{15}
\ee
leads to the total charge,

\bea
Q^{(1)}&:=&\int_{x,\theta}\Sigma^{(1)}\nonumber\\
&\equiv&\frac{1}{4\sqrt{2\lambda}}\int_{x,\theta}V(\Phi)\nonumber\\
&-&\frac{1}{2\sqrt{2\lambda}}\int_{x,\theta}D_-\left(\Phi D_-\Phi\right),\label{16}
\eea
as $D_-^2=0$. The first term on the LHS is expected to yield a constant value, as $V(\Phi,\varepsilon)$
is a infinitesimal deformation away from the periodic potential $V(\Phi)$. The second term is expected to
vanish at the boundary of the superspace, as $\Phi$ is too. This invariably leads us to the {\it conserved} 
charge of the quasi-integrable model, yielding,

\be
\frac{dQ^{(1)}}{dt}=0.\label{17}
\ee
This is not true, in general, for the higher order charges of the model,

\be
\frac{dQ^{(2n+1)}}{dt}=P_{(2n+1)}\neq0,\quad n=1,2,3,
\ee
in general. This makes the system quasi-integrable. It is easy to see that, for $\varepsilon=0$, $X=0$,
which leads to $P_{(2n+1)}=0,\quad n=1,2,3$, and the SG model is recovered.

\section{Summary}
It has been shown that quasi-integrability can be attained in supersymmetric integrable models also, by taking
the super SG model as an explicit example. It is to be mentioned here that the exact periodicity of
the undeformed potential plays a crucial role in this formalism. Further, the extended structure of 
the superspace introduces extended sub-algebras (twisted loop algebras), forming closed vector spaces. 
A further unique aspect, that of non-locality of parameters, is observed.
They yield a few conserved 
charges, as in the case of quasi-integrable non-supersymmetric models, with the others remaining 
non-conserved.

\acknowledgments
The authors are grateful to Professors Luiz. A. Ferreira, Wojtek J. Zakrzewski and Betti
Hartmann  for their encouragement, various
useful discussions and critical reading of the draft.

\end{document}